\def\eq{$\equiv$\/}
\def\l{$\lambda$}
\def\g{$\gamma$}

\def\REF{\par\noindent\hangindent 20pt}
\def\kms{km~s$^{-1}$}

\def\a{$\alpha$}
\def\b{$\beta$}

\def\l{$\lambda$}

\def\ltsima{$\; \buildrel < \over \sim \;$}
\def\simlt{\lower.5ex\hbox{\ltsima}}            
\def\gtsima{$\; \buildrel > \over \sim \;$}
\def\simgt{\lower.5ex\hbox{\gtsima}}            
\def\cm3{cm$^{-3}$}
\documentstyle[12pt]{article}
\begin{document}
\null
\vspace{2cm}
\begin{center}
\baselineskip=24pt
{{\huge \bf Evidence for a Very Broad Line Region in
PG 1138+222}}
\vspace{2truecm}

\centerline{\large\it Paolo Marziani
 \&\ Jack W. Sulentic  \footnote{Visiting Astronomer, Kitt Peak
 National Observatory }}
\vspace{0.5truecm}

{\it \small Department of Physics \&\ Astronomy

University of Alabama

Tuscaloosa, Al 35487-0324

U.S.A.}
\vspace{2cm}

\vspace{4.5cm}

{~~~~~~~~~~~~Received:~~~~~~~~~~~~~~~~~~~~~~~~~~~~~~~~~~Accepted:~~~~~~~~~~~~~~}

\end{center}
\newpage
\baselineskip=24pt
\begin{flushleft}
\vspace{1.2cm}
\centerline{\Large\bf Abstract}

\vspace{1cm}

We present a high S/N spectrum for the quasar PG 1138+222. We detect a very
broad HeII$\lambda$4686 emission line component with twice the FWHM of the
conventional broad line region (BLR) as evidenced by the Balmer lines. The
profile shape and centroid redshift also distinguish this HeII component
from the BLR features. The large ratio of HeII\l 4686 to any analogous
H$\beta$ emission component is an indicator that it arises in a very high
density region ($n_e \simgt 10^{11}-10^{13}$ \cm3). This Helium component
is probably emitted in a Very Broad Line Region (VBLR), where the radiation
field is so strong that the Str\"omgren depth becomes similar to the
geometrical thickness of the emitting clouds. The gas could therefore be
optically thin to the Lyman continuum.

\vspace{1.2truecm}

{\em Subject headings}: Galaxies: nuclei -- Line Formation  -- Line Profiles
-- Quasars: Spectroscopy
\parindent=16pt
\newpage

\section{Introduction}

Photoionization models are fairly successful in reproducing the broad line
spectra of Seyfert galaxies and quasars. Recent attempts to compute the
emergent spectrum of the Broad Line Region (BLR) have led to significant
improvements because a wider range of electron density has been considered
($n_e$ from the ``canonical'' value 10$^9$ up to 10$^{13}$ \cm3 )(Rees,
Netzer \&\ Ferland, 1989; Dumont \&\ Collin--Souffrin, 1990; Ferland et al.
1992). It is still unclear whether density decreases smoothly with distance
from the continuum source or whether the low ionization lines (Balmer
lines, MgII$\lambda$ 2800, FeII emission) and the high ionization lines
(Ly$\alpha$, CIV $\lambda$ 1549, and also HeII $\lambda$ 4686) are emitted
in two different environments (with different $n_e$ and probably different
underlying kinematics). The possibility that LIL and HIL arise in different
environments could  account for the apparent blue shift of the HIL lines
relative to the LIL (Gaskell 1982). Such a difference, for example, arises
naturally in a radiating accretion disk model. The accretion disk might be
the LIL emitter, provided that it is illuminated by a significant amount of
back--scattered continuum radiation (Dumont \&\ Collin--Souffrin, 1990).
One problem with this interpretation is the poor overlap between model
predictions and the observed properties of the emission lines (Sulentic et
al. 1990).

Irrespective of any multi--zoned structure or stratification it is likely
that extreme physical conditions will be encountered at the inner edge of
the BLR. The wings of the Balmer lines and some HIL could arise in a {\em
Very Broad Line Region} (VBLR) whose properties differ markedly from those
of a composite BLR. Although high S/N spectra of Seyfert galaxies often
show very extended and faint wings with FWZI up to 30,000 \kms (see e.g.
Akn 120 \&\ IC 4329A, Marziani, Calvani \&\ Sulentic, 1992) the possibility
of a VBLR has only been discussed for  Mkn 590 (Ferland, Korista \&\
Peterson, 1990) and NGC 5548 (Peterson \&\ Ferland, 1986; Wamsteker et al.
1990) as far as we are aware. The former authors introduced the idea of an
optically thin VBLR for Mkn 590 in order to expain the absence of
variability in the wings of the Balmer lines. A tenfold increase in the
strength of the HeII$\lambda$ 4686 line in NGC 5548, along with a much
smaller increase in H$\beta$\ ,was similarly ascribed to optically thin gas
very close to the continuum source (Peterson \&\ Ferland, 1986). In some
objects (where the OI $\lambda$ 8446 line is observed) the wings of the
Balmer lines are found to be stronger or more extended than those for OI.
This line is believed to arise from gas that is optically thick to the
Lyman continuum (Morris \&\ Ward, 1988). This raises the possibility that
the wings of the Balmer lines arise in an optically thin region. Ferland,
Korista \&\ Peterson (1990) proposed that a ``screen'' of optically thin
gas located between the canonical BLR and the central continuum source
might even reprocess part of the nonthermal continuum and create the {\em
Big Bump}. The standard explanation for the Big Bump involves optically
thick gas, either in an accretion disk, or in dense, filamentary structures
(Lightman \&\ White, 1988).

Little attention has been paid to the HeII$\lambda$4686 feature as a
valuable indicator of the range of properties inside the BLR. This is
probably because this feature is usually contaminated by FeII emission. We
find a very broad HeII$\lambda$4686 component in PG 1138+222. This feature
could only be detected with : (1) high S/N data, (2) no significant
contamination from an underlying galaxy and (3) weak FeII emission between
H$\beta$ and H$\gamma$. Thus PG 1138+222 provides a rare opportunity to
make  a reliable study of the HeII\l 4686\ profile. Our analysis of the
HeII\l 4686 and Balmer line profiles are reported in this paper  along with
a heuristic analysis of the physical condition for gas in the VBLR.

\section{Observations and results}

\subsection{Observations and data reduction}

PG 1138+222 (\eq WAS 26) was observed at Kitt Peak National Observatory on
February 19, 1991, with the Gold Camera attached to the Cassegrain focus of
the 2.1 m telescope. A dispersion of 1.3 \AA/pixel was obtained with a 600
l/mm grating and an 800 $\times$ 800 pixel TI chip. The spectral range
covered was $\approx$ 4500--5500 \AA. The slit width was $\approx$ 2.5
arcsec (yielding a spectral resolution of $\approx$\ 3.2 \AA\ FWHM), while
the seeing was estimated to be $\approx$ 1.5 arcsec. Standard IRAF
procedures were adopted to convert the observed spectrum to linear
wavelength scale. Conversion to absolute flux units was based upon
observations of the star Feige 34.

\subsection{Results}

Figure 1 shows the spectrum of PG 1138+222. The redshift, computed from all
identified narrow emission lines (reported in Table 1) is $z = 0.0632 \pm
0.0004$. Fluxes of the lines, uncorrected for galactic reddening, (Burstein
\&\ Heiles (1984) give $E(B-V)_{gal}=0.0$) are reported in the second
column of Table 1. The underlying continuum was first subtracted with a
third-order spline function in order to estimate the relative contribution
of the broad and narrow components. The  broad components of H$\gamma$,
HeII\l 4686 and H\b were then fitted  with appropriate spline functions.
Results are shown plotted on the same wavelength scale in Figures 2a b and
c respectively. Absolute fluxes of the broad and narrow components are
estimated to be accurate within $\pm$ 20 \%.

The narrow line component ratio H$\gamma$/H$\beta$ $\approx 0.49\pm 0.06
(2\sigma)$ agrees (within the uncertainties of the measurement) with the
value appropriate for the Narrow Line Region (NLR) as computed by Ferland
\&\ Netzer (1983). Several Seyfert galaxies and quasars show a ratio that
is considerably larger. This suggests that it is not always advisable to
infer the reddening from this ratio alone. The lack of other indicators
leads us to assume  negligible internal reddening in PG 1138+222 as a first
approximation.

An interesting result not directly related to the main topic of this paper
is the remarkable strength of the well resolved [OIII]\l 4363 line.
Accepted at a face value, the ratio [OIII]\l\l 4959,5007/[OIII]\l 4363
implies an electron temperature of $T_e \approx 3.5\times 10^{4~\circ}$K
for an electron density $n_e = 10^4$ cm$^{-3}$\ or $T_e \approx 2.8 \times
10^{4~\circ}$K for $n_e = 10^5$ cm$^{-3}$  (Osterbrock, 1988). The ratio
[OIII]\l 4363/H\b$_{NC}$\ $\approx$ 0.44 and the implied $T_e$\ are too
large to be consistent with photoionization. This ratio is too high even
for gas whose abundance is a factor of ten below the solar value and with
very high ionization parameter ($\Gamma \sim 0.01-0.1$, Ferland \&\ Netzer
1983). Heating mechanisms other than photoionization (e.g shocks) are
probably at work in the NLR of PG 1138+222. A complication connected with
the [OIII] line ratio given above involves the observation that \l 4363
profile is nearly twice as broad as \l 5007. The greater FWHM of [OIII]\l
4363 is clearly seen in Figure 2a and  may indicate structure in the NLR.
Gas with $n_e$\ between 10$^6$ and 10$^7$\ \cm3\ may contribute to the
[OIII] \l 4363 emission. A higher density gas would produce a higher
velocity dispersion which might explain the unusual ratio without invoking
shocks.

The principal result of this paper is the detection of a broad component
underlying the narrow HeII\l 4686 emission feature. It is 2.4 times broader
than the mean FWHM for H$\beta$ and H$\gamma$. Similar values of FWZI are
reported in Table 1 for all three lines. However the FWZI measures for the
Balmer lines are affected by the presence of a weak red wing extending up
to $v_r \sim 7000$ \kms, which is not seen in HeII.

The FeII emission in PG1138+222 is weak, but lines from multiplets 27,
42 and 36 are likely contributors to the extended red wings of H$\gamma$\
and H\b: FeII \l 5169 (multiplet 42) is clearly visible at the
red edge of our spectrum. Moreover, the red wing of H\g\ is so weak, that
even FeII emission slightly above noise could reduce or cancel out this
feature. Nevertheless, we believe that the red wing is real because: (1) it
is seen in both the H\b\ and H$\gamma$\ profiles of our spectrum, (2) the
S/N ratio of our spectrum is very high ($\approx$50) and  (3) Osterbrock
\&\ Shaw (1988) published a low resolution spectrum of PG 1138+222 where
the red wing is also visible in H\a.

The Balmer line profiles can be classified as type AR,R (i.e., red
asymmetric profile with redshifted peak) in the scheme proposed by Sulentic
(1989). This type is rather common among quasars and Seyfert galaxies. The
BLR component of H$\beta$\ shows a redshift (at FWHM) of 400$\pm$75 km/s
relative to the narrow line component. The profile of HeII\l 4686, appears
to be markedly different from the Balmer lines. This is clearly seen in the
Figure 2a,b comparison between H$\gamma$\ and HeII which have comparable
fluxes. More quantitatively, if we take the ratio FWHM to FWZI as an
indicator of the ``boxiness'' of a profile, we find 0.25 and 0.28 for
H$\beta_{BC}$\ and H$\gamma_{BC}$\ while for HeII\l 4686, the ratio is
0.67. The peak and FWHM centroid values for the broad HeII component show
the same  redshift as the NLR with an uncertainty of 300 \kms.

Another comparison that emphasizes the difference between the HeII and HI
profiles involves the ratio between HeII\l 4686$_{BC}$\ and H\b $_{BC}$ as
shown in Figure 3.  The ratio was computed after subtraction of the
underlying continuum and NLR components as well as smoothing with a
gaussian filter. The error bars displayed in Fig. 3 represent the
uncertainty due to different, extreme choices of the underlying continuum.
They should be regarded as representative of the maximum error. The ratio
HeII\l 4686$_{BC}$/H\b$_{BC}$\ is about 0.2-- 0.3 in the central part of
the profile (similar to the flux ratio integrated over the entire line
profile). The ratio increases up to a value greater than one at $v_r
\approx - 2000$ \kms, and it remains $\sim 2$ up to $v_r \approx -4000$
\kms, where  both H\b\ and HeII fall below the $3\sigma$\ noise level.
Although the ratio also tends to increase on the red side, its maximum
value is lower, namely $\approx$ 0.8, because of the  red wing on H\b. If
we interpret the red wing as a distinct component or, at least, as arising
in a region different from HeII\l 4686, the values on the blue side can be
assumed to give a more relevant estimate for the ratio. The profile ratio
involving the red side of H$\beta$ will also increase after correction for
FeII contamination.

\subsection{ A comparison of line ratios and the implications}

The average ratio between the equivalent width of HeII\l 4686 and H$\beta$\
is 0.095 $\pm$ 0.10 for the 87 quasars studied by Boroson \&\ Green (1992).
These EW measures include both the broad and narrow line emission. The NLR
contribution is usually small for H$\beta$ while the situation for HeII\l
4686 is less clear.  The ratio of the broad components in PG 1138+222 is
F(HeII\l4686)/F(H\b) $\approx$ 0.33. This value is rather large, and in
fact only 10 quasars in the above sample show F(HeII\l4686)/F(H\b)
$\approx$ 0.20; seven of these show F(HeII\l 4686)/F(H\b) $\approx$\ 0.30.
Only PG 1244+026 (0.51) and PG1534+580 (0.40)  show stronger HeII than PG
1138+222. PG1534 shows a very broad HeII profile similar to PG1138+222 with
little or no contamination from FeII. The contamination from FeII emission
in PG 1244+026 is too strong to make reliable inferences. The ratio for PG
1138+222 is actually more extreme than this comparison suggests. We are
comparing the strengths of composite features that arise in different
regions.  It is likely therefore that the ratio of the broad HeII component
to any analogous one in H$\beta$ {\bf is $\geq$ 2.0}.

A ratio of HeII\l 4686$_{BC}$/H$\beta_{BC}$ $\geq$2-3 is larger than case B
situations ($\approx 0.8$), and much larger than the value expected for the
BLR ($\sim $\ 0.2), where H\b\ is collisionally enhanced by a factor of
$\approx 5$. It is a strong indicator of high density ($n_e \sim 10^{11} -
10^{13}$ \cm3) inside the BLR. Earlier studies (Shuder, 1982; Osterbrock
\&\ Shuder, 1982) based on the HeI\l 5876/H\b\ and H\a/H\b\ profile ratios
showed that the electron density and/or the ionization parameter increased
toward higher radial velocities. Evidence for the VBLR rests with
differences in the width and shape of the broad HeII\l4686 component
compared to the Balmer lines. The HeII\l 4686 and \l 1640 lines in NGC 5548
(Wamsteker et al. 1990) were also found to be substantially broader than
H\b\ . The greater width of the HeII component points toward a velocity
regime nearer the central engine than for the conventional BLR.

\subsection{The formation of HeII lines}

The formation of the HeII\l 4686 line has been thoroughly discussed by
MacAlpine (1981), Grandi (1983), and MacAlpine et al. (1985). MacAlpine
discussed the use of the HeII\l1640/HeII\l4686 ratio as a reddening
indicator for the BLR, and suggested that these lines are likely to have a
ratio close to the recombination value.  The HeII \l 4686 line originates
from the level 4 $\rightarrow$ 3 transition as does  Pa\a. Mechanisms which
can enhance the HeII\l 4686 line over its recombination value are: (a)
collisional excitation from level 2 to 4 and (b) radiative pumping of level
4 due to diffuse Ly$\alpha$ fluorescence, either from Ly\a\ photons
generated inside the cloud or coming from other emission line clouds. The
effectiveness of these processes depends on the He$^+$ population of level
2 ($n_{He^+}(2)$), which depends almost entirely on the diffuse HeII Ly\a\
(i.e. HeII \l 304 photons) field. A reliable estimate of $n_{He^+}(2)$\ is
difficult to compute, because of the  complex transfer properties of the
HeII Ly\a\ line. It is well known that the HeII Ly\a\ photons can be
destroyed through the OIII Bowen florescence mechanism, and by ionization
of neutral hydrogen atoms (e.g. MacAlpine 1981; Netzer, Elitzur \&\
Ferland, 1983).

It is beyond the scope of the present work to attempt a quantitative
prediction which would require detailed photoionization calculations.
Nevertheless, if we assume that a very high density gas ($n_e \sim
10^{11}-10^{13}$ \cm3) is contributing to the emission it is inescapable that
the HeII\l 4686 line will be strongly enhanced over case B. MacAlpine
(1981) computed the number of resonance scatterings by HeII Ly\a\ needed to
sustain a given value of $n_{He^+}(2)$ where the strength of HeII\l 4686 is
due equally to HI Ly\a\ pumping and recombination. At densities $n_e
\sim 10^{11}-10^{13}$\ \cm3, equations (5) and  (10) of MacAlpine (1981)
show that the number of scatterings permits a significant contribution from
collisional excitation to level 4 and, to an even greater extent, from
radiative pumping. The strength of HeII \l 4686 is enhanced by a large
factor over its case B value if the pumping is due to HI Ly\a\ emitted from
other emission line clouds. At densities $> 10^{11}$ \cm3, the strength of
H\b\ is also lowered by collisional deexcitation further contributing to an
increase of the HeII\l4686$_{BC}$/H$\beta_{BC}$ ratio.

Generally speaking, the stronger the He$^+$\ ionizing radiation field, the
stronger the diffuse HeII Ly\a\ field should become, thereby increasing the
population of level 2 in the $He^+$\ atom. Other relevant effects that
enhance HeII\l 4686 over its recombination value depend strongly on
density. The values $\Gamma \sim 0.1$\ and $\sim 0.01$\ for $n_e \sim
10^{11}$ and $\sim 10^{13}$ \cm3\ respectively, should be regarded as the
lower limits necessary to explain our profile ratio. Larger values of the
ionization parameter (i.e. the strength of the radiation field to which the
clouds are exposed if the density is fixed) would create more favourable
conditions for large values of the HeII\l 4686/H\b\ ratio.

The detailed photoionization calculations by Rees, Netzer \&\ Ferland
(1989) indeed show that the ratio HeII\l 4686/H\b\ becomes larger than one
if $n_e \sim 10^{13}$ \cm3 and if the ionization parameter $\Gamma =
Q(H^0)/4 \pi r^2 c n_e$ is greater than 0.01, where Q(H$^0$) is the number
of ionizing photons.  A ratio of around three can also be obtained for $n_e
\sim 10^{11}$ \cm3 and $\Gamma \sim 0.1$\ (cf. Fig. 5 of Ferland et al.
1992). Generally speaking, a stronger radiation field gives a stronger
diffuse HeII\l 304 field, and hence a larger  population of He$^+$ level 2.
Values of $\Gamma$\ larger than those given above (i.e., a stronger
radiation field for densities $\simgt 10^{11}$ \cm3) would create even more
favorable conditions to increase the HeII\l 4686/H\b\ ratio. This problem
deserves more attention from the theoretical point of view. Nevertheless in
light of the previous analysis and the photoionization results, it is
probably appropriate to assume that the density at the innermost edge of
the VBLR is in the range $\sim 10^{11} - 10^{13} $ \cm3. In the following
calculations, we will consider two cases: (1) $n_e \sim 10^{13}$\ \cm3\ and
$\Gamma \sim 0.01$, and  (2) $n_e \sim 10^{11}$\ \cm3\ and  $\Gamma \sim
0.1$, as representative of the physical conditions in the VBLR.

\subsection{Estimate of the VBLR properties for PG 1138+222}

The distance of the VBLR from the central continuum source can be written
as $R_{in} \approx 1.58 \times 10^{16} (\Gamma n_e)_{11}^{-1/2}$ cm. The
number of ionizing photons Q(H$^0$) needed to compute $\Gamma$\ was
estimated by extrapolating the optical continuum up to $h\nu = 30$ keV
assuming an ionizing continuum of the form given by Netzer (1990). We are
unaware of any  far UV or X--ray observations that might help us here.

The filling factor $f_f$\ of the total HeII emitting region was estimated
writing the luminosity of the HeII\l 4686 line as:
\begin{equation} L(HeII\lambda 4686) =
\int n_e n_{He^{++}}
\alpha_B^{eff}(HeII\lambda 4686) h \nu 4 \pi r^2 dr f_f,
\end{equation}
where $\alpha_B^{eff}(HeII\lambda 4686)$\ is the effective recombination
coefficient (case B) for HeII and the integral is to be evaluated
from $R_{in}$. We used $H_0 = 100$\ \kms Mpc$^{-1}$ throughout the paper.
The electron density was assumed to scale as $n_e = n_{e0}(r/R_{in})^{-2}$
\cm3. We considered $n_{e0} = 10^{11}$ and $10^{13}$ \cm3
(as in model B of Rees, Netzer \&\ Ferland, 1989). The recombination
coefficient is
$\alpha_B^{eff}(HeII\lambda 4686) \approx 1.43\times10^{-13}$
cm$^3$s$^{-1}$
at $T_e \approx 2\times 10^4~^\circ$ K (Osterbrock, 1988).
If emission is confined in a shell of depth $0.1 R_{in}$,
and if  $f_f$\ does not depend on r, then $f_f \sim 2.6\times 10^{-6}
n_{e0,12}^{-2} R_{in,16}^{-3}$.

The covering factor  (in the case of spherical geometry) is $f_c = R_{in}
n_e f_f/ 3 N_c \approx 0.034 R_{in,16} n_{e,12} f_{f,-6} N_{c,23}$,
where $N_c$\ is the column density. A broad upper limit to the the covering
factor can be obtained from the relation
\begin{equation}
L(HeII\lambda 4686) = h\nu [ \alpha_B^{eff}
(HeII\lambda 4686)/ \alpha_B^{eff} (He^{++} \rightarrow He^+)] f_c Q(He^+),
\end{equation}
valid if the gas is optically thick for photons with energy higher
than 54.4 eV. The value of $f_c$\ for the HeII emitting gas should thus be
$f_c \simlt 0.29
Q_{54}^{-1}(He^+)$\ ($n_e \sim 10^{13}$ \cm3), and  $0.39
Q_{54}^{-1}(He^+)$\ ($n_e \sim 10^{11}$ \cm3), where the number of He$^+$\
photons is in units of 10$^{54}$. Ferland, Korista \&\ Peterson (1990)
suggest that the optically thin gas in Mkn 590 is distributed in a screen
that completely covers the central source. In their view, the central
source of power--law radiation is surrounded by ionized and partially
transparent gas. However employing a {\em bare} power--law $f_\nu \propto
\nu^{-\alpha}$\ with $\alpha = 1.0$\ to compute Q($H^0$) we obtain an upper
limit to $f_c \simlt 0.5-0.6$ for PG 1138+222. This suggests that there is
no need to have complete coverage of the continuum source of photons above
the He$^+$ threshold in order to account for the observed HeII\l 4686
luminosity.

Strong variations in HeII\l 4686 coupled with a weaker variation in the
Balmer lines suggested that the HeII emitting gas in NGC 5548
(Peterson \&\ Ferland, 1986) was optically thin to the Lyman continuum. The
same conclusion probably holds for PG 1138+222. The depth of the
He$^{++}$\ Str\"omgren sphere is $R_{He^{++}} = 3
\Gamma(He^{+})c/n_{He^{++}} \alpha_B^{eff}(He^{++} \rightarrow He^+)$,
where $\alpha_B^{eff}(He^{++} \rightarrow He^+)$\ is the total effective
recombination coefficient, and $\Gamma(He^+)$ is the ionization parameter
for He$^+$, that is $\Gamma(He^+) = Q(He^+)/ 4 \pi n_e c R_{in}^2$\ (which
is $\sim$\ 0.1 $\Gamma(H^0)$\ with our choice of the continuum). The
number of $He^+$\ photons was again estimated assuming that
the UV -- X-ray continuum had the form tabulated by Netzer (1990). Assuming
$n_{He^{++}} \sim 0.1 n_e$, and using the recombination coefficient
computed by Hummer \&\ Storey (1987), we find R$_{He^{++}} \sim
6.5\times10^{10}$ cm ($n_e \sim 10^{11}$ \cm3) and R$_{He^{++}} \sim
4\times10^{7}$ cm  ($n_e \sim 10^{13}$ \cm3).

The Str\"omgren radius for H$^+$ is analogously $R_{H^+} = 3 \Gamma(H^0)
c/n_e \alpha_B^{eff}(H^+ \rightarrow H^0)$, and corresponds to $\sim
6\times 10^{8}$  and $\sim 6\times 10^{11}$\ cm for $n_e = 10^{13}$ and
$10^{11}$\ \cm3\ respectively. If $n_e \sim 10^{11}$ \cm3, the canonical
value for the column density of active nuclei, $N_c \sim 7\times 10^{22}$
cm$^{-2}$\ is obtained at $R \approx 7\times 10^{11}$ cm.  This suggests
that the HeII emitting gas is most probably optically thin to the Lyman
continuum. However, if $n_e \sim 10^{13}$ \cm3, $\Gamma $\ would have to be
$\sim 0.1$\ in order for the Str\"omgren depth to become comparable to the
geometrical thickness of the clouds. Furthermore, the canonical value of
$N_c$\ given above refers to neutral gas. It is probably a reliable
estimate of the total cloud $N_c$\ under standard BLR conditions, where the
Str\"omgren radius should be much smaller than the geometrical thickness of
the clouds. The simplest choice was to also adopt this value of the column
density for the VBLR.

\subsection{Further considerations}

Great care should be taken in using HeII\l 1640/HeII\l 4686 ratio as a
reddening indicator. If the line profile of HeII\l 4686 is significantly
broader than that of H\b, as is the case for PG 1138+222 and as  may be the
case for most HeII--strong objects, emission in the line wings could give a
non-negligible contribution to the total flux. As outlined above, there are
reasons to believe that emission in the wings comes from high density gas
in a VBLR exposed to an extremely strong radiation field. In these
conditions, the HeII\l 4686 is expected to be strongly enhanced with
respect to case B of recombination theory and to remain optically thin, while
the optical depth in the HeII\l 1640 line could become larger than 1
(Ferland et al. 1992).  Differences in the  enhancement of HeII\l 4686 and
\l 1640 may depend primarly on the relative contributions of clouds of
different densities (exposed to different radiation field intensities),
whose distribution inside the emitting region is unknown.

There is a kinematic difference between our VBLR evidence and that from
Ferland et al. (1990). Our broad HeII\l4686 component is centered at or
near the systemic (as indicated by the NLR) velocity of the quasar. Their
broad H$\beta$ component shows a significant redshift
$\Delta$V$\approx$10$^3$km/s. The difference in redshift between the HeII
feature and the underlying Balmer features suggests that both cannot arise
in the same kind of optically thin region. A similar comment applies to
Wamsteker et al (1990) where the (VBLR) HeII \l 4686 feature in NGC 5548
shows a 2$\times$ 10$^3$km/s blueshift. We would only expect a
gravitational redshift in the simplest static model for a VBLR. The
differing  velocity results suggest that obscuration and/or radial motion
effects are present in the VBLR.

\section{Conclusion}

The introduction of a VBLR is not motivated simply for taxonomical
convenience. The heuristic value of the VBLR stems from the physical
conditions of the gas located closest to the central continuum source: as
the distance decreases, the depth of the Str\"omgren radius also increases,
making the gas optically thin to the HI Lyman continuum. We suggest that
the large value of the ratio HeII\l 4686$_{BC}/H\beta_{BC} > 1$ can be
explained if the VBLR is made up of dense clouds with $n_e \sim
10^{11}-10^{13}$ \cm3. If the above assumptions are true the HeII emission
arises in  region that may not reprocess most of the continuum emission.
Thus, it could not account for the UV emission from the ``big bump''
(Ferland et al. 1990). This filtering environment might not  be the ``warm
absorber'' suggested by X-ray observations (Yaqoob, Warwick \&\ Pounds
1989).

Other points remain to be clarified. The first is whether the quasars with
strong HeII emission should be regarded as the high-end tail of a
distribution or as a  different population of objects. It is also unclear
whether the large spread in the ratio HeII\l 4686$_{BC}$/H\b$_{BC}$\ ($\sim
0 - 0.88$\ among the Seyfert 1 galaxies studied by Osterbrock, 1977)
reflects inclination effects, the distribution of matter inside the BLR,
or, less probably, intrinsic differences in the ionizing continuum. Since
the VBLR is the innermost line emitting region, it could be hidden if the
central engine is not seen nearly face--on.

\section{References}

\REF

\REF
Boroson T., Green R.: 1992, ApJ 80,109
\REF
Burstein D., Heyles C., 1984, ApJS 54, 33
\REF
Dumont A.--M., Collin--Souffrin S.: 1990, A\& A 229, 313
\REF
Ferland G.J., Netzer H., 1983, ApJ 264, 105
\REF
Ferland G.J., Korista K.T., Peterson B.M.: 1990, ApJ 363, L121
\REF
Ferland G.J., Peterson B.M., Horne K., Welsh W.F., Nahar S.N., 1992,
ApJ 387, 95
\REF
Gaskell, M. 1982, ApJ 263, 79
\REF
Grandi S., 1983, ApJ 268, 591
\REF
Hummer D.G.,  Storey P., 1987, MNRAS
\REF
Lightman A.P., White T.R., 1988, ApJ 335, 57
\REF
MacAlpine G.M., 1981, ApJ 251, 465
\REF
MacAlpine G.M., Davidson K., Gull T.R., Wu C.C., 1985, ApJ 294, 147
\REF
Marziani P., Calvani M., Sulentic J.W., 1992, ApJ 393, 658
\REF
Morris S.L., Ward M.J., 1989, ApJ 340, 723
\REF
Netzer H., Elitzur Ferland G.J., 1983, ApJ 299, 752
\REF
Osterbrock D.E., 1977, ApJ 215, 733
\REF
Osterbrock, 1988, Astrophysics of Gaseous Nebulae and Active Galactic
Nuclei (Mill Valley: Univ. Science Books)
\REF
Osterbrock D.E., Shaw R.A, 1988, ApJ 327, 89
\REF
Osterbrock D.E., Shuder J. M., 1982, ApJS 49, 149
\REF
Netzer H., 1990, in Active Galactic Nuclei, p. 57 (Berlin: Springer).
\REF
Peterson B.M., Ferland G.K., 1986, Nature 324, 345
\REF
Rees M.J., Netzer H., Ferland G.K., 1989 ApJ 347, 640
\REF
Shuder J.M., 1982, ApJ 259, 48
\REF
Sulentic J.W. 1989, ApJ 343, 54
\REF
Sulentic J.W., Calvani M., Marziani, P., Zheng W., 1990, ApJ 355, L15
\REF
Yaqoob T., Warwick R.S., Pounds K.A., 1989 MNRAS 236, 153
\REF
Wamsteker W., et al., 1990, ApJ 354, 446

\section{Figure Caption}

\REF

\REF
Fig.1: Spectrum of PG 1138+222. Horizontal scale is wavelength in \AA;
vertical scale is flux in units of 10$^{-15}$ ergs s$^{-1}$~cm$^{-2}$
\AA$^{-1}$.
\REF
Fig.2: Fits to the broad components and to the underlying continuum
of the strongest lines: (a) H$\gamma$; (b) HeII\l 4686; (c) H\b.
Scale units are as for Fig. 1.
\REF
Fig.3: Profile ratio between HeII\l 4686$_{BC}$\ and H\b$_{BC}$. Horizontal
scale is radial velocity (binned to 225 \kms). The origin has been set at
the average radial velocity of the narrow components of the two lines.

\section{Acknowledgements}

 P.M. acknowledges the financial support provided by Italian Consiglio
Nazionale delle Ricerche for a short post-doctoral period at the University of
Alabama. Extragalactic studies at the University of Alabama are supported
by  NSF/State of Alabama EPSCoR grant RII-8996152.

\end{flushleft}
\end{document}